\documentclass{aa520}
\usepackage{epsfig}
\newcommand\kms{\,km\,s$^{-1}$}
\begin{document}

\title{The radial velocities and physical parameters of HD~553\thanks{Based
      on observations made with the Coud\'e-spectrograph at the 
      2m-RCC-telescope of the National Astronomical
      Observatory Rozhen, Bulgaria.}
     }

\author{R.~Duemmler\inst{1}, I.~Kh.~Iliev\inst{2}, L.~Iliev\inst{3}}

\offprints{L.\ Iliev, liliev@libra.astro.bas.bg}

\institute{Astronomy Division, P.O.~Box 3000, FIN-90014 University of
           Oulu, Finland \and
           National Astronomical Observatory (NAO) Rozhen, and Isaac Newton
           Institute of Chile, Bulgarian Branch, P.O.~Box 136, BG-4700
           Smolyan, Bulgaria \and
           Institute of Astronomy, Bulgarian Academy of Sciences, and Isaac Newton
           Institute of Chile, Bulgarian Branch, Tsarigradsko
           Shosse Blvd.~72, BG-1184 Sofia, Bulgaria}

\date{Received date; Accepted date}

\authorrunning{R.~Duemmler et al.}
\titlerunning{HD~553: radial velocities and physical parameters}

\abstract{\object{HD~553} was discovered as an eclipsing binary by Hipparcos.
          Here, we present the first radial velocity curve for this system. It
          is found, that HD~553 is a double-lined spectroscopic binary. Despite
          the large luminosity difference, the two components of this system
          are of very similar mass. The
          primary, a K0-giant, fills a large fraction of its Roche-lobe. The
          secondary is, despite its very similar mass, still  a late-type dwarf.
          The radial velocity curve allows to
          constrain several stellar and system parameters.
          \keywords{stars: individual: HD~553 -- binaries: spectroscopic --
                    binaries: eclipsing -- stars: late-type -- stars: starspots
                    -- techniques: radial velocities}
       }
\maketitle

\section{Introduction}
\label{S-intro}

Additional to the astrometric parameters of positions, proper motions and
trigonometric parallaxes, the highly successful Hipparcos mission also provided
photometric information for the target stars. Thus, a large number of new
variables has been discovered by Hipparcos. Among these is HD~553
(=HIP~834; $\alpha_{2000}=00^{\rm h}\,10^{\rm m}\,10\fs4647,\ 
 \delta_{2000}=+64^{\circ}\,38\arcmin\,48\farcs174$; V$\approx8\fm2$).
Kazarovets et al.~(\cite{IBVS99}) subsequently classified the system as an
 eclipsing binary of the
$\beta$~Lyr type (i.e.\ two minima are present and there is no constant part
due to the ellipticity effect) and gave it the name \object{V\,741~Cas}.
Strassmeier et al.~(\cite{Strassmeier00}) identified HD~553 as a single-lined
spectroscopic binary and included it into their list of potential
targets for Doppler-imaging from their survey of stellar activity indicators.

The Hipparcos catalogue (ESA \cite{ESA97}) 
provides the following information about this system:
\begin{itemize}
\item V=$8\fm11$, B--V=$1\fm032\pm0\fm016$, V--I=$1\fm00\pm0\fm01$
\item $\pi=4.74\pm 0.83$ mas
\item period $P=9.0576\pm0.0008$ days, T$_0$=2448504.0610 as time of zero phase
      (from fits to the Tycho-lightcurve).
\end{itemize}

HD~553 is an ideal target for Doppler imaging. At observatories north
of latitude $26^{\circ}$, like the NAO Rozhen, Bulgaria
($\varphi\approx 42^{\circ}$), it is circumpolar,
allowing for in principle continuous observations. This would allow to follow
the spot evolution on any timescale (longer than the rotation period) 
without the often 1 yr timegap between
subsequent maps for other Doppler imaging targets. 
Assuming that the orbital and rotational axes of the eclipsing binary
are aligned, the inclination of the rotational axis is high allowing
to observe most of the stellar surface.  It is bright enough to allow for good
signal-to-noise spectra within a reasonably short time. Because the eclipse, when the primary is
in the back, is far from total we can also resolve most of the stellar surface
during the eclipse. Its projected rotational velocity $v\,\sin\,i=42.5$\kms\ (see
Sect.\ \ref{S-RVmeas}) is
high enough to allow for reliable Doppler imaging even with a moderate
resolution of about 30\,000, as obtainable with the Coud\'e spectrograph at
Rozhen. Finally, the period of 9 days is long enough that one good spectrum per
observing night is enough to provide a good phase coverage, and thus the object
can easily be included in any spectroscopic night's object list. On the other hand,
the period is short enough to allow the data for a complete map to be collected
within the typical 1--2 week spectroscopic observing runs.

Therefore, we became interested in this object and started an observing campaign.
The first step for reliable Doppler imaging is to find a good radial velocity
curve. The problem of Doppler imaging is ill-posed and has
a lot of free parameters. Thus, any additional constraints that can be put prior
to Doppler imaging is of great value. Since Doppler imaging relies on
distortions in the line profiles caused by the surface structures, it is
important to remove any other distortion of the line profiles, in particular any
shift of the lines with respect to the same lines at other phases. In principle,
the radial velocity shifts from the orbital motions could be removed prior to
Doppler imaging by simply cross-correlating the spectra. However, the spots on
the stellar surface introduce systematic errors in the shifts obtained in this
way. It is therefore useful to determine the radial velocity based on a large
time scale; each individual radial velocity will still be influenced by the line
profile distortions, but the limited lifetime of the spots ensures that spectra
at the same phases taken a long time apart average out these distortions. Also,
any reliable determination of the orbital parameters is based on a good value
for the orbital period (all other parameters more or less correlate with the
orbital period), which becomes more and more accurate the longer the time base
of the radial velocity curve.

Here, we report on the first radial velocity curve of the HD~553 system
based on 20 high-accuracy radial
velocities obtained over the time base of 2 years, and the conclusions about
the system and stellar parameters that can be drawn from it.

\section{Observations and reductions}
\label{S-obs}

The observing material consists of 20 spectra obtained with the Coud\'e
spectrograph at the 2m-RCC-telescope of the National Astronomical Observatory
Rozhen, Bulgaria. The data were collected in 4 observing runs in
October/November 1999, April and October 2000, and September/October 2001.

The grating 632/$22\fdg3$ and an entrance slit width of 300 $\mu$m were used,
resulting in a resolution of about 30\,000. The spectra were recorded with the
Photometrics AT200 camera on a 1024$\times$1024 CCD with pixel size
24$\times$24\,$\mu$m$^2$,
yielding a spectral range of about 10 nm for each spectrum.
In different seasons, different spectral regions have been observed: in 1999,
the region around the \ion{Ca}{i} line at 643.9 nm, in 2000 and 2001, the region
around the \ion{Fe}{i} line at 617.3 nm have been observed; additionally, in
2001, another region, that around a group of telluric lines at 630 nm, was
observed immediately before or after the \ion{Fe}{i}-region, which allows an
independent check of the wavelength scale via the atmospheric lines and of the
radial velocities (RVs) via the comparison of the RVs from the different
regions at virtually the same time. The average signal-to-noise ratio of the
spectra is 136.

Additional to the spectra of HD~553, several spectra of the RV
standard \object{$\beta$~Gem}
have been obtained in each observing run,
except for Oct 2000, when we observed the RV standard \object{$\alpha$~Cas}.
Both stars are classified as K0\,III, and are therefore suitable templates for
HD~553.

The information for each spectrum is given in Table~ \ref{T-RV}\footnote{
Table \ref{T-RV} is also available in electronic form at the CDS via anonymous
ftp to cdsarc.u-strasbg.fr (130.79.128.5) or via http://cdsweb.u-strasbg.fr/Abstract.html}.

Each observed stellar spectrum is accompanied by a pair of comparison spectra
(one taken before, the other after the stellar exposure) and a number of
flatfield images. In several nights, a large number of bias-images was obtained,
which were found to be unchanging within an observing run; they were averaged,
then subtracted from each image of the same run.

The reduction of the material has been done using the spectrum reduction package
4A (Ilyin \cite{Ilyin00}), and consists of the steps bias image subtraction,
flatfielding, scattered light subtraction, optimal spectrum extraction,
wavelength calibration.

The wavelength
calibration is based on the comparison spectrum images consisting of Th-Ar emission
lines. Each Th-Ar-image consists of a pair of comparison spectra: one is above,
the other below the area covered by the stellar spectrum, but overlapping with
it. This allows to take into account a slope of the lines of constant wavelength
across the stellar spectrum. The observation of a comparison image before and
another one after the stellar exposure allows to take into account a possible
systematic drift of the pixel coordinate system along the wavelength axis.
Thus, the final dispersion curve of each stellar image is based on a total of
4 Th-Ar-comparison spectra, and is interpolated to the central position of the
stellar spectrum and the time of mid-exposure of the stellar spectrum.
The dispersion curve is approximated by a polynomial of degree 3 (except for
Sept/Oct 2001, where a degree of 5 was necessary) and has typically an accuracy
of 0.02--0.04\kms.

After the wavelength calibration, the spectra are transformed into heliocentric
wavelengths and normalized to the apparent continuum which is a polynomial of
degree 4 approximating the high points of the stellar spectrum.

It is worth mentioning that 4A estimates the errors of each pixel in the original
image (after bias subtraction) assuming Poissonian noise and propagates these
errors through each step of the reduction procedure. In this way, reliable errors
are available for each pixel in the final spectrum in both flux and wavelength
resulting also in individual errors for each RV.

\section{Radial velocity measurements}
\label{S-RVmeas}

Radial velocities were determined for each spectrum by cross-correlation with the
average of all spectra of the radial velocity standard observed during the same
observing run. For this, the spectra of the standards were
artificially spun up to different values of the projected rotational velocity
$v\,\sin\,i$, and the spectrum which maximizes the cross-correlation peak height
gives the best value of $v\,\sin\,i$. This was done for both standards $\beta$~Gem
and $\alpha$~Cas and for all spectra of HD~553. The result is
$v\,\sin\,i=42.5\pm0.5$\kms, independent of which standard is used, independent
of the spectral region and of the phase in the binary orbit. The latter allows
the conclusion that the ellipticity of the primary in the binary (the giant)
is not very pronounced.

Before cross-correlating the artificially broadened spectra of the templates
with the spectra of HD~553 they were corrected for the radial velocity of the
corresponding standard star.
For $\beta$~Gem, RV=$+(3.3\pm0.1)$\kms\ according to the Astronomical
Almanach for 2002 (\cite{AstAlm01}) and Stefanik et al.\ (\cite{Stefanik99}).
For $\alpha$~Cas, in recent publications several different RVs are given,
casting some doubt on its suitability as a radial velocity standard:
RV=$-(3.9\pm0.1)$\kms\ according to the Astronomical Almanach for 2002 (\cite{AstAlm01}),
RV=$-(4.3\pm0.2)$\kms\ according to Udry et al.\ (\cite{Udry99}), and
RV=$-(4.79\pm0.04)$\kms\ according to de~Medeiros \& Mayor (\cite{Medeiros99}).
Our investigation (see Sect.~\ref{S-RVcurve}) shows that the value by de~Medeiros \& Mayor
(\cite{Medeiros99}) is best.
Note, that by observing the standard stars in the same run as the spectra of
HD~553 (usually immediately before or after HD~553) and reducing them in exactly
the same way, then moving them to RV=0 using the values from the literature,
the RVs of HD~553 are on the system of the literature velocities of the standards,
without any contribution from some systematic velocity zero-points (which could
be different in different runs) (see also Sect.~\ref{S-RVcurve}).

\begin{figure}
\centerline{\resizebox{9cm}{!}{\includegraphics[angle=90]{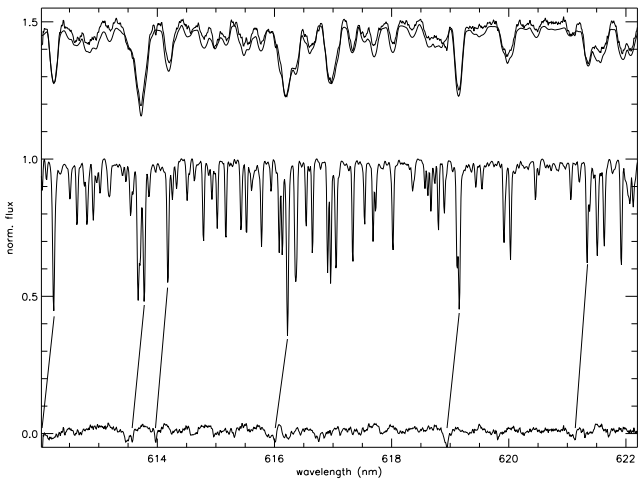}}}
\caption{\label{F-second}
         The spectrum of HD~553 (thick line, top) compared to the spun-up
         spectrum of $\beta$~Gem ($v\,\sin\,i=42.5$\kms; thin line), both
         shifted up by 0.5. At the bottom is the difference between these two
         spectra, showing some weak, narrow absorption lines of the secondary.
         The central spectrum is the original spectrum
         of $\beta$~Gem, used to measure the RV of the secondary, which is here
         --101\kms\ with respect to the primary (HJD 2452185.5178).
         The spectra of HD~553 are in the restframe
         of the primary (determined from a preliminary orbit fit, based only on
         the primary's RVs).
        }
\end{figure}

We also searched for signs of the secondary in the spectra. Indeed, a faint
spectrum displaying narrow
lines moving in antiphase to the strong, broad lines of the giant were found,
after the spectrum of $\beta$~Gem had been subtracted from that of HD~553.
The RVs from these lines were measured by cross-correlation with the original
(not spun-up) spectrum of the RV standard of that run. RVs of the secondary
are not measureable when the secondary is eclipsed. Figure~\ref{F-second}
shows the spectrum of HD~553 in the 617.3\,nm region and identifies some lines
of the secondary.
It can be seen that the continuum of the broadened $\beta$~Gem spectrum
tends to be below that of HD~553. The reason is the imperfect continuum
normalization for HD~553, which is due to the very low signal-to-noise
and to the fact that its (pseudo-)continuum is only defined by a very
few regions without broad, blended features.
However, the difference spectrum is mostly flat so that
RV measurements should not be influenced and no attempt has been made to
correct for this. Note that some ``emission'' features appear in the difference
spectrum which result from an imperfect match of some of the features in the
artificially broadened $\beta$~Gem spectrum and in the HD~553 spectrum. Yet,
the lines of the secondary are clearly visible and measureable.

All the strong lines in the $\beta$~Gem spectrum are represented
in the spectrum of the secondary; several weaker features are also identifiable.
The lines are narrow, which can be seen e.g.\ with the clearly
resolved double feature near 613.5\,nm.
This
 indicates a much lower $v\,\sin\,i$ of the secondary; together with the
low luminosity and the presence of the same strong lines as in the spectrum
of the late giant $\beta$~Gem, this makes the secondary a late-type dwarf.
Therefore, we decided to use the unbroadened spectum of $\beta$~Gem as a
template for the RV measurements of the secondary. Earlier experience has
shown that a good RV of $\beta$~Gem can be measured with the Sun as a template.
So, a mismatch of the spectral types of $\beta$~Gem and of the secondary should
have only a small effect on the final RV of the secondary.

Unfortunately, the lack of proper standards, the comparitively low
signal-to-noise ratio in the spectrum of the secondary, and the also visible
systematics resulting from the not perfect match of the broadened $\beta$~Gem
spectrum and the HD~553 spectrum make it impossible to analyse the spectrum
of the secondary in more detail.

\begin{table}
\caption
      {\label{T-RV}
       The radial velocities of HD~553. The columns are:
       The heliocentric Julian date (HJD--2450000) of the mid-exposure,
       the approximate centre of the
       spectral region in nm, the signal-to-noise ratio (SNR) of the spectrum,
       the radial velocities and errors of the primary and secondary, respectively,
       in \kms.
      }
\vspace*{-2mm}
\begin{center}
\begin{tabular}{ccrr@{$\pm$}lr@{$\pm$}l}
\cline{1-7}
HJD--2450000 & $\lambda_c$ & SNR & \multicolumn{2}{c}{RV$_1$} & \multicolumn{2}{c}{RV$_2$}\\
\cline{1-7}
1478.6010 & 644 &  76 &    17.26 & 0.38 &  -97.45 & 0.77 \\
1481.3821 & 644 & 162 &   -88.47 & 0.38 &   11.87 & 0.66 \\
1482.4879 & 644 & 157 &  -102.36 & 0.26 &   25.36 & 0.69 \\
1483.3475 & 644 & 189 &   -89.04 & 0.23 &   12.81 & 0.55 \\
1484.3707 & 644 & 151 &   -51.71 & 0.20 & \multicolumn{2}{c}{---} \\
1485.3671 & 644 & 174 &    -9.59 & 0.23 &  -70.83 & 0.54 \\
1648.5779 & 617 &  73 &    -3.36 & 0.40 &  -74.37 & 0.69 \\
1652.3007 & 617 &  59 &   -40.38 & 0.32 & \multicolumn{2}{c}{---} \\
1831.3489 & 617 & 160 &    23.94 & 0.35 & -103.73 & 0.53 \\
1832.2459 & 617 & 121 &     8.41 & 0.36 &  -89.52 & 0.68 \\
1832.6138 & 617 & 182 &    -3.91 & 0.34 &  -77.18 & 0.50 \\
1833.3567 & 617 & 160 &   -33.31 & 0.29 & \multicolumn{2}{c}{---} \\
1834.3547 & 617 & 169 &   -74.61 & 0.36 &   -0.42 & 0.53 \\
2181.4355 & 630 & 120 &   -73.98 & 0.44 &   -3.81 & 1.63 \\
2181.4664 & 617 & 125 &   -71.79 & 0.26 &   -3.30 & 0.54 \\
2182.5348 & 630 & 126 &   -28.18 & 0.33 & \multicolumn{2}{c}{---} \\
2182.5675 & 617 & 147 &   -26.87 & 0.19 & \multicolumn{2}{c}{---} \\
2183.5393 & 630 & 102 &    10.27 & 0.45 &  -89.74 & 1.18 \\
2185.4905 & 630 & 145 &    11.20 & 0.36 &  -90.93 & 1.06 \\
2185.5178 & 617 & 132 &    10.69 & 0.27 &  -90.48 & 0.37 \\
\cline{1-7}
\end{tabular}
\end{center}
\end{table}

All RVs are given in Table~\ref{T-RV} together with their errors as determined by
4A. These take into account the errors of the pixels in the spectra in the
overlapped region of both the programme and the standard spectrum as well as the
height and curvature of the cross-correlation maximum (for details, see Ilyin
\cite{Ilyin00}).

\section{Orbital solution}
\label{S-RVcurve}

\begin{table}
\caption{\label{T-orbpar}
         The orbital parameters of the double-lined spectroscopic binary HD~553.
         Errors of the parameters are based on 10\,000 bootstrap runs.
         The primary derived parameter is $P_{\rm rest}$, the period in the
         restframe of the binary system, i.e.\ the observed period corrected for
         the Doppler-effect caused by $\gamma$. All dynamical derived parameters
         are based on $P_{\rm rest}$, while the epochs $T_0$, $T_{\rm conj}$ are
         in the observer's system, i.e.\ based on $P_{\rm obs}$. $T_0$ is the
         time of maximum RV for the primary (the giant), $T_{\rm conj}$ is the
         conjugation time with the primary in the back. $\chi^2$ is the reduced
         $\chi^2$ of the fit, i.e.\ the weighted sum of squared deviations,
         divided by the number of degrees of freedom. $\sigma$ is the mean
         standard deviation of a single RV with average weight, $\sigma_1$ and
         $\sigma_2$ are the same for the primary and secondary, respectively.
        }
\begin{tabular}{lr@{$\pm$}l}
\cline{1-3}
parameter & \multicolumn{2}{c}{\ }\\
\cline{1-3}
$K_1$  (\kms)               & 63.32        & 0.19\\
$K_2$  (\kms)               & 66.04        & 0.41\\
$P_{\rm obs}$ (days)        & 9.05997      & 0.00011\\
$\gamma$ (\kms)             & --38.93      & 0.12\\
$T_0$ (HJD)                 & 2451813.0877 & 0.0036\\[0.3cm]
$P_{\rm rest}$ (days)       & 9.06115      & 0.00011\\
$q=m_2/m_1$                 & 0.9589       & 0.0066\\
$a_1\,\sin\,i$ ($R_{\sun}$) & 11.336       & 0.034\\
$a_2\,\sin\,i$ ($R_{\sun}$) & 11.822       & 0.073\\
$m_1\,\sin\,i$ ($M_{\sun}$) & 1.037        & 0.013\\
$m_2\,\sin\,i$ ($M_{\sun}$) & 0.995        & 0.009\\
$T_{\rm conj}$ (HJD)        & 2451815.3527 & 0.0035\\[0.3cm]
$\chi^2$                    & \multicolumn{2}{c}{2.14}\\
$\sigma$ (\kms)             & \multicolumn{2}{c}{0.56}\\
$\sigma_1$ (\kms)           & \multicolumn{2}{c}{0.59}\\
$\sigma_2$ (\kms)           & \multicolumn{2}{c}{0.97}\\
\cline{1-3}
\end{tabular}
\end{table}

The radial velocities in Table~\ref{T-RV} were used in a double-lined orbit fit.
A few details about the fit routine can be found in Duemmler et 
al.\ (\cite{Duemmler97}).
The weights of the individual RVs were obtained as the inverse variances, based
on the RV errors. However, it turned out, that the primary and secondary had
very different values of $\chi^2$; therefore, the weights of the secondary
were reduced by a common factor $\chi^2_1/ \chi^2_2=0.108$.

\begin{figure}
\centerline{\resizebox{9cm}{!}{\includegraphics[angle=90]{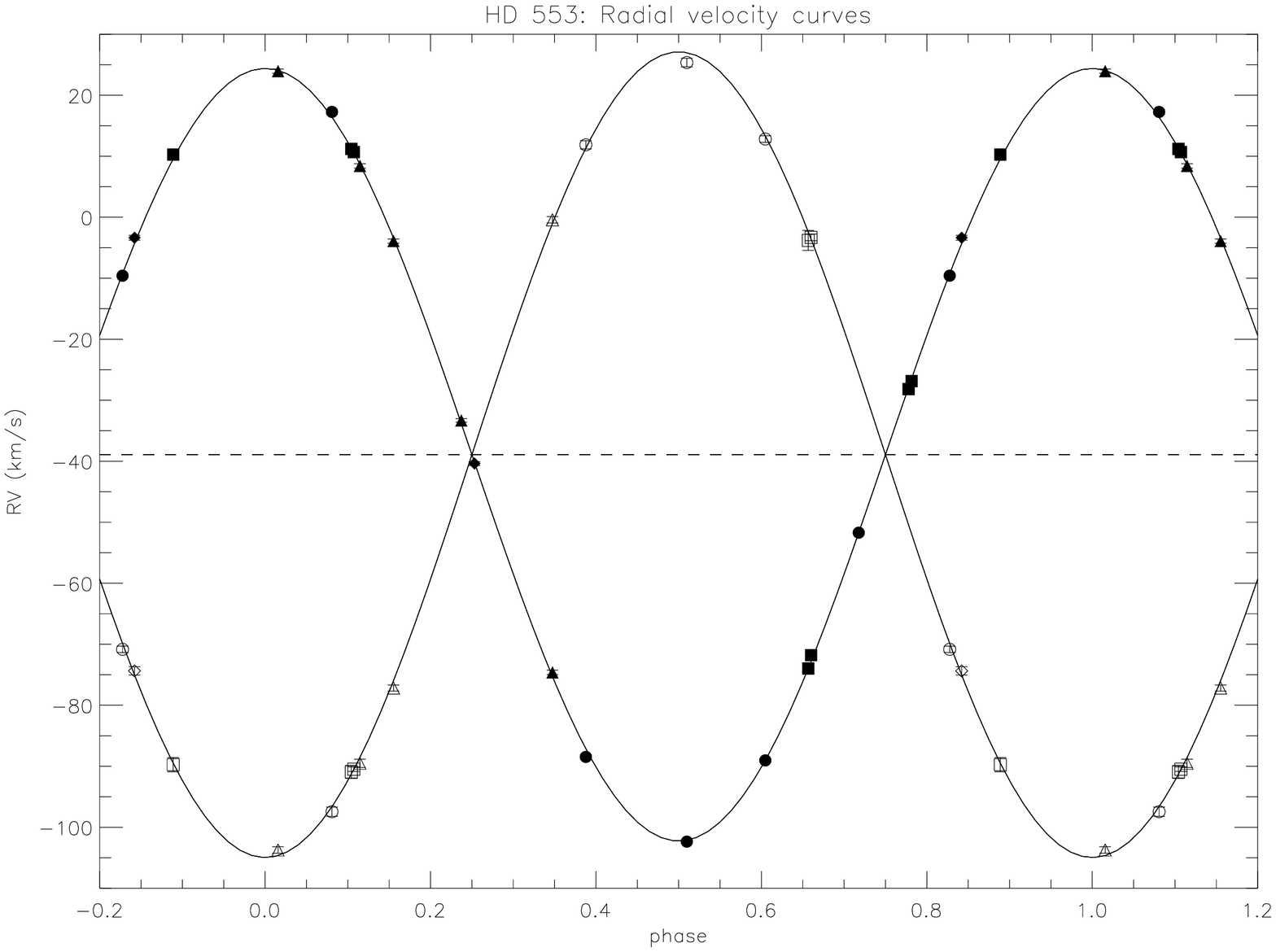}}}
\centerline{\resizebox{9cm}{!}{\includegraphics[angle=90]{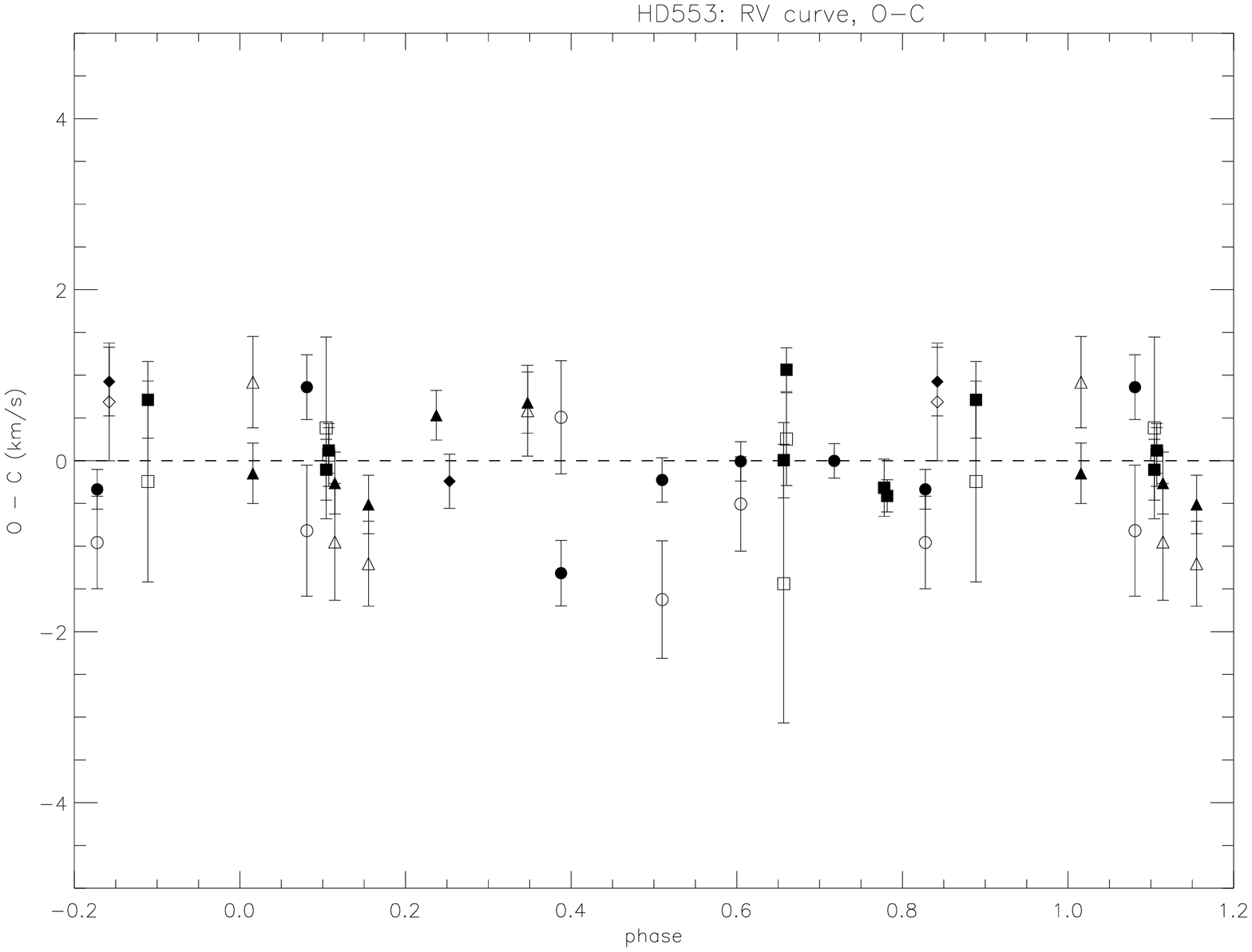}}}
\caption{\label{F-orb}
         {\it top:}
         The radial velocity curves of the two stars in the HD~553 system.
         Filled symbols are used for the primary (the giant), open symbols for
         the secondary; circles denote the data for Oct/Nov 99, diamonds those
         for Apr 00, triangles those for Oct 00, and squares those for
         Sep/Oct 01. Errorbars are typically smaller than the symbol size.\newline
         {\it bottom:}
         O--C vs.\ phase.
        }
\end{figure}

While the large ratio of the $\chi^2$ values of the primary and secondary 
have
 shown that the errors for the secondary given in Table~\ref{T-RV} are 
clearly
underestimated, the overall reduced $\chi^2$ of the fit of 2.1 is quite good.
Although the formal probability of exceeding a value of 2.0 is just 1\% for 15
degrees
of freedom, experience with many fits has shown that, even for larger numbers
of RVs with {\bf smaller} errors, values of 2--3 for $\chi^2$ are not uncommon. However,
larger deviation than expected from the errors may also result from stellar
activity: starspots distort the line profiles, and cause the RV to deviate
systematically. Therefore, we have investigated the O--C values as a function
of phase and observing run, but we do not find any systematic (see
Fig.\ \ref{F-orb}). One should, however, keep in mind that 20 measurements
is still a rather small statistics.

The errors of
the fit-parameters were determined by 10\,000 bootstrap runs
(see Duemmler et al.\ \cite{Duemmler97} and, e.g., Efron \& Tibshirani
\cite{ET93}). To each RV computed from the orbital solution we
add a randomly chosen value taken from the O--C values.
Since on average low weighted
RVs tend to have larger O--C, each O--C is accompanied by its original weight;
the choice of O--C values is done with replacement, i.e.\ the same O--C can
be chosen several times while others are not used. With this artificial data
set a new orbital fit is performed, leading to a slightly different set of
parameters than the fit to the original data. The error of a 
parameter is given by the standard deviation of this parameter from all
the bootstrap samples. The resulting
orbital and derived parameters are presented in Table~\ref{T-orbpar}. The
orbit is plotted in Fig.~\ref{F-orb}.

Strassmeier et al.\ (\cite{Strassmeier00}) have also measured two RVs
for HD~553. While these two RVs increase the time span of the RV curve by another
year, the large errors compared to our RVs lead to very small weights in the fit.
The result is identical to the fit presented in Table~\ref{T-orbpar}, except
that the errors of the parameters even increased slightly. We therefore decided
to ignore the Strassmeier et al.\ measurements.

We also attempted to fit an eccentric orbit.
 The result $e=0.0016\pm 0.0023$ with
$\omega=350\fdg7\pm 0\fdg1$ clearly shows that the orbit is circular. The
Lucy-Sweeney $F$-test (Lucy \& Sweeney \cite{Lucy71}, Lucy \cite{Lucy89})
gives a 96.2\,\% probability for a circular orbit.

Another fit allowed for different $\gamma$ velocities for the 4 observing runs,
while forcing all other parameters to be the same for the 4 data sets. That
means that we are looking for systematic RV errors between the different runs,
or for systematic changes of the $\gamma$ velocity from run to run, which could
be caused by a third body. Within the errors, the orbital parameters do not
change compared to those given in Table~\ref{T-orbpar}; the results for $\gamma$
are:
\begin{eqnarray}
{\rm Oct/Nov\ 1999}\hspace{0.3cm} \gamma_1&=&-39.27\pm 0.19\nonumber\\
{\rm Apr\ 2000}\hspace{0.3cm}     \gamma_2&=&-38.62\pm 0.34\nonumber\\
{\rm Oct\ 2000}\hspace{0.3cm}     \gamma_3&=&-38.73\pm 0.25\\
{\rm Sep/Oct\ 2001}\hspace{0.3cm} \gamma_4&=&-38.75\pm 0.26.\nonumber
\end{eqnarray}
The largest difference, $\gamma_1-\gamma_2$, is only 1.7\,$\sigma$, i.e.\ there
is no indication for any significant change of $\gamma$ over the two-year
interval or for a significant error in the velocity zero points of the different
runs. It is noteworthy that $\gamma_3$ is based on RVs measured with $\alpha$~Cas
as standard, for whose RV we adopted the value given by de Medeiros \& Mayor
(\cite{Medeiros99}). Would we have adopted the other values given in the
literature, $\gamma_3$ would deviate by 0.9\kms (Astronomical Almanach for 2002
(\cite{AstAlm01})) and 0.5\kms (Udry et al.\ \cite{Udry99}), respectively; it is
clear, that the value by de Medeiros \& Mayor (\cite{Medeiros99}) is best, at
least at the time of our measurements, Oct 2000.

It is remarkable that the orbital parameters in Table~\ref{T-orbpar} show a system
with almost identical masses for the two stars, although one of the stars is
clearly evolved.

It should be mentioned that our period $P_{\rm obs}$ is longer than both the period
given in the Hipparcos catalogue (ESA \cite{ESA97}; $P=9\fd0576\pm0\fd0008$) and the
period given
by Strassmeier et al.\ (\cite{Strassmeier00}; $P=9\fd0580$).
While Strassmeier et al.'s period is consistent with the one given by Hipparcos,
our period is longer by $3\sigma$. However,
 since the Hipparcos
period is based on a curve fit, whose shape is not specified, we feel that our period
is more reliable.

\section{Conclusions and discussion}
\label{S-Conclusion}

We have found the Hipparcos-detected eclipsing binary HD~553 to be a
double-lined spectroscopic binary with a giant as primary and a faint
companion of almost equal mass. From 20 high-quality RVs collected over the
course of 2 years, the
radial velocity curve was derived, leading to an improvement of the binary
period and the derivation of other orbital and physical parameters.

If we assume that the primary is in synchronised rotation (i.e.\ the system is
old and the primary has been a giant for a long time), the rotation and orbital
axes should also be aligned, i.e.\ $P_{\rm rot}=P_{\rm orb}$, for
which the period $P_{\rm rest}$ in the rest of frame of the system should be
used and
$i_{\rm rot}=i_{\rm orb}=i$. Then, the projected radius is given by $v\,\sin\,i$
via
\begin{equation}
\label{E-radvsini}
R_1\,\sin\,i=\frac{P\,v\,\sin\,i}{2\,\pi}=(7.608\pm 0.090)\,R_{\sun},
\end{equation}
with $v\,\sin\,i=(42.5\pm0.5)$\kms\ as determined in Sect.~\ref{S-RVmeas}.

The assumption of aligned axes and synchronisation
of rotation and orbit could, however, be wrong.
While it is true that the system is old for the primary to become a
giant by now, so possibly the axes are aligned, the periods may be different.
During the expansion of the primary towards the giant branch the rotation period
should increase due to angular momentum conservation. 
A longer rotation period would increase the radius given above.
It would also imply that
the primary's expansion happened quite recently, so that the tidal forces were not
able to speed up again the primary's rotation. This is highly likely, because the
secondary is still a dwarf, despite it having nearly the same mass as the primary.

Therefore, it is good to compare this lower limit for the radius with the radius that
can be obtained from the photometric properties. According to Strassmeier
et al.\ (\cite{Strassmeier00}), V=$8\fm11$, B--V=$1\fm03$, T$_{\rm eff}=4790$\,K; and
according to the Hipparcos catalogue (ESA \cite{ESA97}), $\pi=4.74$\,mas, corresponding to a distance of
211\,pc. The V-magnitude could be too faint, if the star has been covered by spots at
the time of the measurements and B--V could be too red; additionally, the interstellar
extinction ($A_{\rm V},\ E_{\rm B-V}$) is unknown. This means, that the resulting absolute
magnitude $M_{\rm V}=1\fm49$ is an upper limit. With BC=$0\fm43$ (read off from
Fig.\ 10.16, p.\ 197, in Gray \cite{Gray92}), the upper limit for the bolometric
absolute magnitude
becomes $M_{\rm bol}=1\fm03$, and the lower limit for the luminosity 29\,$L_{\sun}$,
leading to a lower limit of 7.85\,$R_{\sun}$ for the radius of the primary. This value
is consistent with the lower limit from Eq.\ (\ref{E-radvsini}).
 Based on the good subtraction of the $\beta$~Gem spectrum from that of
HD~553, we could repeat the calculation using the temparature of $\beta$~Gem:
T$_{\rm eff}=4865$\,K (Drake \& Smith \cite{Drake91}); this leads to
(7.61\,$R_{\sun}$), identical to the lower limit in Eq.\ (\ref{E-radvsini}).

While both results are lower
limits, it would be quite a coincidence, if the difference of the rotation and orbital
periods and the projection factor $\sin\,i$ on the one hand and the spot coverage and
interstellar absorption on the other lead to the same change in the radius. It is more
likely, that $\sin\,i_{\rm rot}\approx \sin\,i_{\rm orb}\approx 1$,
$P_{\rm rot}\approx P_{\rm orb}$, V$\approx$V$_{\rm unspotted}$, and $A_{\rm V}\approx 0$
(despite a galactic latitude of just 2$^{\circ}$).

The effective radius of the Roche-lobe of the primary, $R_{\rm RL}$,
i.e.\ the radius of the sphere having the
same volume as the Roche-lobe, is another quantity with which to compare the
lower limit for the primary's radius. According to Eggleton~(\cite{Eggleton83})
\begin{eqnarray}
R_{\rm RL}\,\sin\,i&=&a\,\sin\,i\ \frac{0.49\,Q^{2/3}}{0.6\,Q^{2/3}+\ln(1+Q^{1/3})}\nonumber\\
                   &=&(8.859\pm0.034)\,R_{\sun}
\end{eqnarray}
with the inverse mass ratio $Q=m_1/m_2=1.0429\pm 0.0072$.

The fact that the Roche-lobe radius is only about 1\,$R_{\sun}$ larger than the radius of
the primary shows that the assumption of synchronisation cannot be grossly wrong: If the
rotation period were longer than $10\fd6$ (in the restframe of the primary), the star
would overfill its Roche-lobe and most likely form a shell around the whole system, which
seems not to be the case. Thus, the deviation from synchronicity cannot be larger than
16\,\%.

If we consider synchronous rotation
as valid, the star fills a considerable fraction (63.3\,\% of the volume)
of its Roche-lobe. Still, since the star is nearly 40\,\% smaller than the
Roche lobe, we would not expect at present a large deviation from sphericity
(see also Sect.\ \ref{S-RVmeas}, where it was found that $v\,\sin\,i$ does not
depend on phase).
According to Gray (\cite{Gray92}, App.\ B), a K0\,III
star has on average a radius of $R({\rm K0\,III})=11\,R_{\sun}$. Either the primary
in HD~553 is still expanding, in which case it will soon (in evolutionary terms)
fill its Roche-lobe and start losing mass towards the secondary, or its inclination
makes $R\,\sin\,i=7.6\,R_{\sun}$ so much smaller than the typical radius, in
which case the inclination would be about 44$^{\circ}$. Since the system is
eclipsing, the latter possibility seems highly unlikely; given the similar
masses of the two components, but the very different luminosities, the former
possibility seems to be very plausible. Thus, we expect HD~553 to become soon
a semidetached system.

Finally, the small errors in the radial velocity curve
presented here, based on the long time span of 2 years, should allow a good
transformation of future spectra into the restframe of the primary for
surface imaging. However, many RS~CVn stars show persistent active
longitudes with activity cycles lasting several years 
(Berdyugina \& Tuominen \cite{Berdyug98}). Thus, the relatively small
amount of statistics, based on only 20 measurements over the comparatively
short time span of 2 years may not be enough to ensure that all systematic
effects from spots on the radial velocity curve are removed yet.
Future spectra will therefore also be used to further improve the orbital
parameters.

\begin{acknowledgements}
We thank the referee, Dr.\ Holger Lehmann, for his careful reading of the
original version of the paper and for his useful questions and suggestions,
which helped to improve the paper.
A large part of this work was financially supported by the
Academy of Finland and the Bulgarian Academy of Sciences through travel grants
for RD and LI, for which we are grateful.
We made use of the SIMBAD database, maintained at the Centre de Donn\'ees
Stellaire (CDS) in Strasbourgh, France.
\end{acknowledgements}

\end{document}